\documentstyle[seceq,epsf,wrapft]{ptptex}

\title{Large-$S$ approach to chiral phases in frustrated spin chains}

\author{
Alexei K. {\sc Kolezhuk}\footnote{E-mail address: 
kolezhuk@itp.uni-hannover.de}
}

\inst{
Institute of Magnetism, National Academy of Sciences and Ministry of
Education,
36(B) Vernadskii avenue, 03142 Kiev, Ukraine\footnote{Permanent address}\\ 
and Institut f\"ur Theoretische Physik, Universit\"at Hannover,\\
Appelstra{\ss}e 2, 30167 Hannover, Germany}

\recdate{
\today
}

\abst{ A large-$S$ approach to the description of chiral
phases in frustrated antiferromagnetic spin-$S$ chains with easy-plane
anisotropy is proposed.  The approach is valid in the vicinity of the classical
Lifshitz point $j={1\over4}$, where $j$ is the relative strength of
the frustrating next-nearest-neighbor coupling. The effective theory
turns out to be the nonlinear $\sigma$-model with the additional
fourth-order term in spatial derivatives.  The topological term
persists for $j>{1\over4}$, which leads to disappearance of the chiral
gapped phase for half-integer $S$, in accordance with the recent
numerical results.  }

\begin{document}

\maketitle

\section{Introduction}

Chirally ordered phases in 
frustrated quantum spin chains have attracted considerable attention recently
\cite{NersGogEss98,Kab+99,AllenSen00,AligBatEss00,%
Hik+00,K00,Nishiyama00,Hik+01,LJA01}.  
Nersesyan
{\em et al.}\cite{NersGogEss98} have studied the
antiferromagnetic $S={1\over2}$ chain with easy-plane anisotropy and
frustrating next-nearest-neighbor (NNN) coupling, described by the
Hamiltonian
\begin{equation} 
\label{ham} 
\widehat{H}=\sum_{n}\{ ({\mathbf S}_{n}{\mathbf S}_{n+1})_{\Delta} +
j({\mathbf S}_{n}{\mathbf S}_{n+2})_{\Delta},\quad
({\mathbf S}_{1}{\mathbf S}_{2})_{\Delta}\equiv
S_{1}^{x}S_{2}^{x} + S_{1}^{y}S_{2}^{y} +\Delta S_{1}^{z}S_{2}^{z}\,,
\end{equation}
where 
${\mathbf S}_{n}$ denotes the spin operator at the $n$-th site,
 $j>0$ is the relative strength of
the NNN coupling, and $0<\Delta<1$ is the dipolar easy-plane
anisotropy.
In the limit $j\gg 1$, performing the Abelian bosonization 
and subsequently using RG and mean-field arguments, they have
predicted the existence of a new {\em gapless\/} phase
with a broken parity, which is characterized by the nonzero
value of the {\em vector chirality}
${\mathbf \kappa}_{n} = \langle({\mathbf S}_{n}\times{\mathbf
S}_{n+1}) \rangle$.
This type of ordering does not break the U(1) in-plane rotation
symmetry, it only breaks the discrete parity symmetry and thus is in
principle perfectly allowed in one dimension, the idea 
probably first realized by Chubukov. \cite{Chubukov91}
Except having the chiral order, 
this phase is characterized by the power-law decaying incommensurate
in-plane spin correlations of the form 
$\langle S^{+}_{0} S^{-}_{n}\rangle \propto  n^{-\eta}e^{iQn}$,
where $Q$ is very close to $\pi$ in the limit $j\gg1$, and
$\eta={1\over4}$ for $S={1\over2}$.\cite{NersGogEss98}

Early attempts \cite{Kab+99,AligBatEss00} to find this {\em chiral
gapless} phase in numerical calculations for $S={1\over2}$ were
unsuccessful. At the same time, to much of surprise, DMRG studies for
frustrated $S=1$ chain \cite{Kab+99,Hik+00} have shown the presence of
{\em two} different types of chiral phases, {\em gapped} and {\em
gapless}. Later, chiral phases were numerically found for
$S={1\over2}$, \cite{Nishiyama00,Hik+01} as well as for $S={3\over2}$
and $S=2$. \cite{Hik+01}

For general $S$, the
appearance of two chiral phases was  explained  
with the help of the large-$S$ mapping to a helimagnet, and the
qualitative form of the phase diagram for large $S$ was
given.\cite{K00} However, the way of mapping to a helimagnet used in
Ref. \cite{K00} neglects the presence of the topological term and thus
is in fact valid only for integer $S$, when the topological term is
ineffective.  

Another theoretical approach using bosonization \cite{LJA01} suggests
that the phase diagram for integer and half-integer
$S$ should be very similar, with the only difference that the Haldane
phase gets replaced by the dimerized phase in the case of half-integer
$S$. This is, however, in  contradiction with the recent numerical
results \cite{Hik+01} indicating that the chiral gapped phase is
absent for half-integer $S$. 

In the present paper I show that the difference between integer and
half-integer $S$, well known for small $j$, \cite{reviews2} persists
also for the chiral region of the phase diagram, which leads to
disappearance of the chiral gapped phase and also changes dramatically
the mechanisms of destabilization of the chiral gapless phase.

\section{Modified nonlinear sigma model}

Consider the model (\ref{ham}) for general (large) spin $S$. One can
pass to the spin coherent states in
a usual manner, as described, e.g., in Ref.\ \cite{reviews1}. 
The Berry phase for a single spin at the site $i$ can be expressed
through the unit vector $\vec{n}_{i}$ parametrizing the coherent state:
$\Phi_{i}= S\int dt\,{ \partial_{t}\vec{n}_{i}\cdot
(\vec{n}_{i}\times\vec{e}_{i}) \over 1+\vec{n}_{i}\cdot\vec{e}_{i}}$,
where $\vec{e}$ is an arbitrary unit vector. 
If we choose $\vec{e}_{1}=-\vec{n}_{2}$ and
$\vec{e}_{2}=-\vec{n}_{1}$,
the sum of Berry phases
for two neighboring spins at sites $1$ and $2$ can be written easily
in a compact form : 
$\Phi_{12}= S {\vec{n}_{1}\times\vec{n}_{2}\over 1-
\vec{n}_{1}\cdot \vec{n}_{2}}\cdot \partial_{t}(\vec{n}_{2}-\vec{n}_{1})$.
Further, we select the uniform and staggered components of the
magnetization, putting
$\vec{n}_{i}= \vec{m}_{i} +(-1)^{i}\vec{l}_{i}$.
When we are in the vicinity of the classical Lifshitz point
$j={1\over4}$, both uniform and staggered magnetization vary slowly in
space, so that we can pass to the continuum approximation, assuming as
usual that $m\ll l$. 
The
effective action is readily obtained in the following form:
\begin{eqnarray}
\label{Aml}
{\cal A}&=&2\pi S Q +S\int dx \int
dt\,\vec{m}(\vec{l}\times\partial_{t}\vec{l})\\ 
&-& S^{2}\int dx \int dt\, \Big\{ 
{3\over4}(1-\Delta)l_{z}^{2}+2f_{\alpha}m_{\alpha}^{2}
-2\varepsilon h_{\alpha}(\partial_{x}l_{\alpha})^{2}
+{1\over8} h_{\alpha}(\partial^{2}_{xx}l_{\alpha})^{2}\Big\}\,,\nonumber 
\end{eqnarray}
where $Q={1\over4\pi}\int dt\int dx \vec{l}\cdot (\partial_{x}\vec{l}\times
\partial_{t}\vec{l})$ is the topological charge,
and we use the notation
\begin{equation}
 \varepsilon=j-1/4,\quad
f_{x,y}=h_{x,y}=1,\quad f_{z}=(3+5\Delta)/8,\quad h_{z}=\Delta.
\end{equation}
For the sake of clarity, we have set the Planck constant and the lattice
constant to unity.
Making Taylor expansions of the fields, one has to take into account
the derivatives of $\vec{l}$ up to the fourth order, since, as one can
see from (\ref{Aml}), the contribution of
the second order comes with the prefactor $\varepsilon$ which becomes
negative in the region $j>{1\over4}$ we are interested in. Note also
that we do not assume $\Delta$ to be close to $1$ and thus have to
keep terms like $(1-\Delta)$ multiplied by $m_{z}^{2}$ or
$(\partial_{x}l_{z})^{2}$ etc.
The uniform part $\vec{m}$ can be integrated out, and
after passing to the imaginary ``time'' $\tau=2iSt$ one obtains the
following effective Euclidean action, valid in the vicinity of
the Lifshitz point:
\begin{eqnarray}
\label{Anlsm}
{\cal A}_{E} &=&  {1\over 2g_{0}} \int dx \int d\tau \Big\{
{ 1\over f_{\alpha}}(\vec{l}\times\partial_{\tau}\vec{l})_{\alpha}^{2}
-4\varepsilon\,h_{\alpha}(\partial_{x}l_{\alpha})^{2}
+ {1\over4}\,h_{\alpha}\,(\partial^{2}_{xx}l_{\alpha})^{2}\nonumber\\
&+&{3\over2} (1-\Delta)(l_{z})^{2} \Big\} + i 2\pi S Q,
\end{eqnarray}
where the bare coupling constant $g_{0}=2/S$. It is easy to see that
for $\Delta\to 1$ and $j\to 0$ the action (\ref{Anlsm}) gives the
well-known expression for the isotropic Heisenberg chain (the
fourth-order derivatives become irrelevant in this limit).

\section{Mapping to a helimagnet}

Now I will show that the action (\ref{Anlsm}) can be further mapped to
a helimagnet, giving the results very similar to those obtained with
the help of the ansatz of Ref.\ \cite{K00}. Passing to the angular
variables $l_{x}+il_{y}=\sin\theta e^{i\varphi}$, $l_{z}=\cos\theta$,
one may notice that for $\Delta\not=1$ the field $\theta$ is massive,
and thus can be integrated out. Putting $\theta={\pi\over2}+\vartheta$
and expanding in $\vartheta$, one can rewrite the action as
\begin{eqnarray}
\label{heli1}
{\cal A}_{E}&=&{\cal A}[\vartheta]
+{1\over 2g_{0}}\int dx \int d\tau
\Big\{ {1\over
f_{z}}(\partial_{\tau}\varphi)^{2}[1+(f_{z}-2)\vartheta^{2}]\nonumber\\
&+&(1-\vartheta^{2})\big[ -4\varepsilon(\partial_{x}\varphi)^{2} 
+{1\over4}(\partial_{x}\varphi)^{4}+{1\over4}(\partial^{2}_{xx}\varphi)^{2}
\big]\Big\} +i 2\pi S Q,\\
{\cal A}[\vartheta]&=& {1\over 2g_{0}}\int dx \int d\tau \Big\{
(\partial_{\tau}\vartheta)^{2}-4\varepsilon \Delta
(\partial_{x}\vartheta)^{2}
+{1\over4}\Delta(\partial^{2}_{xx}\vartheta)^{2}
-{3\over2}(1-\Delta)\vartheta^{2}
\Big\}\nonumber
\end{eqnarray}

Further, introducing instead of $\vartheta$ a new variable $\xi$ by putting
$\vartheta={1\over2}(\xi e^{i\lambda
x}+\xi^{*}e^{-i\lambda x})$,
where $\xi$ is assumed to be small and smoothly varying, one can
obtain the action ${\cal A}[\xi]$ where now only quadratic terms in
$\xi$ have to be kept. In order to kill terms of the type
$\xi\partial_{x}\xi^{*}-\xi^{*}\partial_{x}\xi$, it is necessary to
set $\lambda=\lambda_{0}$, where 
$\lambda_{0}=\sqrt{8\varepsilon}$
 is
the classical pitch of the helix in the limit $\varepsilon\ll 1$.
The action now takes the form
\begin{eqnarray}
\label{heli2}
{\cal A}_{E}&=&
\int dx \int d\tau 
\Big\{   {1\over
2\widetilde{g_{0}}}
(\partial_{\tau}\varphi)^{2}\big[1+{1\over2}(1+Z)|\xi|^{2}\big]
+{1\over 2g_{0}}\big[1-{1\over2}|\xi|^{2}\big] V[\varphi]
\Big\} \nonumber\\
&+&{\widetilde{E}\over g_{0}}\int dx \int d\widetilde{\tau} \Big\{
|\partial_{\widetilde{\tau}}\xi|^{2} + |\partial_{x}\xi|^{2}
+m_{0}^{2}|\xi|^{2}\Big\} +i 2\pi S Q,
\end{eqnarray}
where the following notation is used:
\begin{eqnarray}
\label{not}
&&\widetilde{g_{0}}= g_{0}(1-Z),\quad Z={5\over8}(1-\Delta),\quad
 m_{0}^{2}={3(1-\Delta)\over 16 \varepsilon\Delta },\quad 
\widetilde{E}=\sqrt{{\varepsilon \Delta\over2}}\nonumber\\
&&\widetilde{\tau}=\tau\sqrt{8\varepsilon \Delta},\quad
V[\varphi]={1\over4}\big[(\partial_{x}\varphi)^{2}-8\varepsilon\big]^{2} 
+{1\over4}(\partial^{2}_{xx}\varphi)^{2}.
\end{eqnarray}
Applying the standard Polyakov-type RG, one obtains finally the
effective action of the planar helimagnet, which depends only on 
in-plane angle $\varphi$:
\begin{eqnarray}
\label{Aheli}
&&{\cal A}[\varphi]={1\over 2 T_{\rm eff}} \int dx \int dy \Big\{
(\partial_{y}\varphi)^{2}+V[\varphi]\Big\} +i2\pi S Q,\quad
T_{\rm eff}\equiv\sqrt{g\widetilde{g}},\quad
y= \tau\sqrt{\widetilde{g}/ g}\nonumber\\
&& g={g_{0} \over 
1-\displaystyle{g_{0}\over 8\pi \widetilde{E}}
\ln(1+\Lambda_{0}^{2}/m_{0}^{2})},
\quad
\widetilde{g}={\widetilde{g_{0}} \over 
1-\displaystyle{\widetilde{g_{0}}\over 8\pi \widetilde{E}}{1+Z \over 1- Z}
\ln(1+\Lambda_{0}^{2}/m_{0}^{2})
}, 
\end{eqnarray}
here $\Lambda_{0}=\pi$ is the momentum cutoff on the lattice. The
problem is thus mapped to the classical helimagnet in two dimensions
at the finite temperature $T_{\rm eff}$.

One may notice that we have kept the topological term in
(\ref{Aheli}), although formally we don't have the right to do that
after mapping to the planar model. Keeping this term is, however,
important for the following discussion, if one considers singular
configurations like vortices: In the vortex core the planar mapping
becomes invalid, since the angle $\theta$ deviates there strongly from
$\pi/2$.

\section{Phase transitions}

The action (\ref{Aheli}) is very similar to that derived in Ref.\
\cite{K00} (although slightly different in detail), and we
will see that it yields exactly the same results concerning the
transition lines.
%
%

\begin{wrapfigure}[22]{r}{65mm}
\epsfxsize=64mm
\centerline{\epsfbox{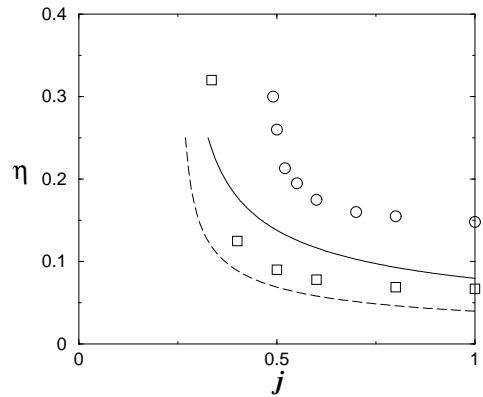}}
\caption{\label{fig:eta12} Behavior of the critical exponent $\eta$ of
the in-plane correlations in the chiral gapless phase
(see Eq.\ (\ref{eta})) as a function of $j$. Solid and dashed lines
are the results of present approach for $S=1$ and $S=2$, respectively,
and circles and squares are numerical results for $S=1$ and $S=2$
taken from Ref.\ \cite{Hik+01}.}
\end{wrapfigure}

The transition from gapless chiral to gapped chiral phase is
determined by the unbinding of vortices existing on the background of
a state with a certain chirality. 
The critical ``temperature'' of this transition can be obtained by
rewriting the action (\ref{Aheli}) in terms of deviation from the
ground state with a certain chirality, i.e., setting
$\varphi=\lambda_{0}x+\phi$. Then one may neglect the higher-order
terms in derivatives of $\phi$ and get the classical $XY$ action of
the form
\[
{\cal A}={1\over 2T_{\rm eff}}\int\!\!\int\!\! dx\,dy\,\big\{
(\partial_{y}\phi)^{2}+\lambda_{0}^{2}(\partial_{x}\phi)^{2}\big\}, 
\] 
from which the critical $T_{\rm eff}$ is given by the equation
\begin{equation} 
\label{TKT} 
T_{\rm eff}^{KT}=(\pi/2)\lambda_{0}
\end{equation}
For $\varepsilon\to 0$ and finite (not very small) anisotropy
$1-\Delta$ one has $g\simeq g_{0}$, $\widetilde{g}\simeq
\widetilde{g_{0}}$, and (\ref{TKT}) translates into
\begin{equation} 
\label{eKT1} 
\varepsilon_{KT}=(3+5\Delta)/4\pi^{2}S^{2}
\end{equation}
exactly the result given in Ref.\ \cite{K00}. In the other limit, when
$1-\Delta\to0$ and $\varepsilon$ is small but finite, 
one may neglect the difference between $g$ and $\widetilde{g}$, which
yields the following equation for the transition line:
\begin{equation} 
\label{eKT2} 
(1-\Delta)_{KT}=(16/3)\varepsilon \Lambda_{0}^{2}
\exp\{-2(\pi S\sqrt{2\varepsilon}-2)\},
\end{equation} 
again the result coinciding with that previously obtained.\cite{K00}

The in-plane correlation function in the KT (gapless chiral) phase can be
readily obtained and has the power-law form
\begin{equation} 
\label{eta}
\langle S^{+}(x) S^{-}(0)\rangle \propto  x^{-\eta}\,
e^{i(\pi\pm\lambda_{0})x},\quad
 \eta=T_{\rm eff}/(2\pi\lambda_{0})
\to {1\over8\pi S}\sqrt{{3+5\Delta\over\varepsilon}},\;\; \varepsilon\to0
\,.
\end{equation}
The critical exponent $\eta$ increases when one approaches the
transition; 
this behavior of $\eta$ is in qualitative agreement with the numerical
results\cite{Hik+01}, see Fig.\ \ref{fig:eta12}. One should mention,
however, that the estimated numerical values of $\eta$ are not
universal at the transition boundary, while the picture advocated here
would imply the universal KT value of $\eta={1\over4}$. Further
studies are necessary to clarify this point. The large-$j$ behavior of
$\eta$, which is not accessible by the present approach, was studied
by Lecheminant et al. within the bosonization framework,\cite{LJA01}
and the predicted value $\eta(j\to\infty)=1/8S$ is in a good agreement
with the numerical data.\cite{Hik+01}

\begin{figure}
\epsfxsize=85mm
\centerline{\epsfbox{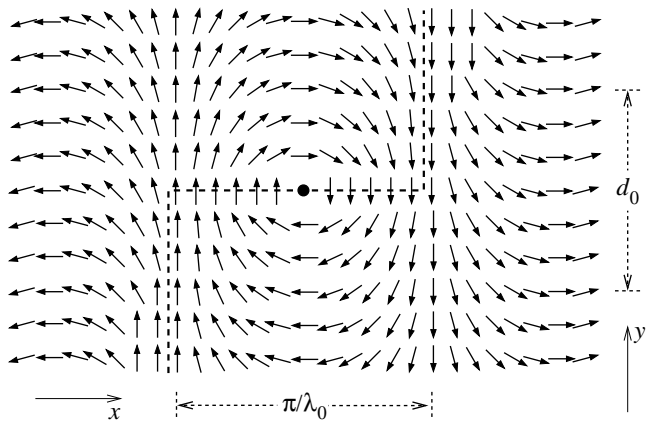}}
\caption{\label{fig:fluctDW} 
A schematic view of the elementary  ``fluctuation'' in a
chiral domain wall: an instanton connecting two domain wall states
separated by the minimum possible distance $\pi/\lambda_{0}$
(displacement by one ``quantum''). The instanton necessarily includes
a vortex-like topological defect and thus has the topological charge
$Q=\pm {1\over2}$. The dashed line indicates the position of the
domain wall.} 
\end{figure}

The temperature of the transition between chiral and non-chiral phases
can be estimated in a ``solid-on-solid'' approximation \cite{MHZ77} 
by considering the fluctuations of a chiral domain
wall.\cite{K00} Any such fluctuation requires formation of  topological
defects with nonzero vorticity, as shown in Fig.\
\ref{fig:fluctDW}. 
Such field configurations, which we will call ``bound
vortices'',  are instantons connecting
states with different positions of the domain wall in space, and the
corresponding change in the position is quantized in multiples of
$\pi/\lambda_{0}$. The free energy $\sigma$ of the domain wall per
unit length in the $y$ direction
can be written as 
\begin{equation} 
\label{freedw1}
\sigma =E_{DW}-{T_{\rm eff}\over L}\ln {\cal Z},\quad
{\cal Z}=\sum_{N=0}^{N_{0}}\int_{y_{N-1}}^{L} dy_{N}\cdots 
\int_{0}^{y_{2}} dy_{1} 
\sum_{\{n_{i}\not=0\}}
e^{\displaystyle -{E_{bv}\over T_{\rm eff}} \sum_{i=1}^{N}|n_{i}|},
\end{equation}
where $L$ is the length of the domain wall, $E_{bv}$ is the energy of
the bound vortex, $E_{DW}$ is the energy of the domain wall (per unit length),
$N_{0}=L/d_{0}$, where
$d_{0}$ is the minimal possible distance in $y$ direction between
successive fluctuations, and the positions $y_{i}$ of fluctuations are
subject to the constraint $|y_{i}-y_{i+1}|\le d_{0}$. The multiple
integral in (\ref{freedw1}) is easily calculated to be equal to
$(L-Nd_{0})^{N}/N!$ by making the substitution $\zeta_{i}=y_{i}-(i-1)d_{0}$,
and after applying the Stirling formula one finally gets
\begin{equation} 
\label{freeDW} 
\sigma=E_{DW}-(T/d_{0})\ln\Big\{
1+d_{0}\big(\mbox{cotanh}(E_{bv}/2T_{\rm eff})-1\big)\Big\}
\end{equation}

\begin{wrapfigure}{r}{66mm}
\epsfxsize=65mm
\centerline{\epsfbox{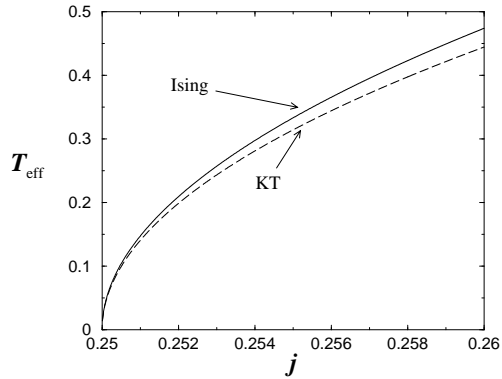}}
\caption{\label{fig:ikt} The effective critical ``temperatures''
$T_{\rm eff}$ of the Ising (chiral --- non-chiral) and Kosterlitz-Thouless
(chiral gapless --- chiral gapped) transitions,
according to Eqs.\ (\ref{TKT}) and (\ref{freeDW}). One can see that
the region of existence of the chiral gapped phase is very narrow.} 
\end{wrapfigure}

We cannot calculate the energy of the bound vortex $E_{bv}$ and the
scale $d_{0}$ analytically. Assuming that bound and free vortex
solutions have similar structure apart from the core region, and
cutting the free vortex solution at the distance $\pi/\lambda_{0}$,
one may estimate $E_{bv}\simeq \pi\lambda_{0}\ln(\pi/\lambda_{0})$ and
$d_{0}=\pi/\lambda_{0}^{2}$. Then the equation $\sigma=0$ can be
solved numerically, and the corresponding solution $T_{\rm eff}^{I}$
as a function of $j$ for $j\to 1/4$ is shown in Fig.\ \ref{fig:ikt}
together with the corresponding critical temperature of the
Kosterlitz-Thouless transition. One can see that the two temperatures
are very close to each other, but $T^{I}$ is slightly higher than
$T^{KT}$.

Thus, for integer $S$ we obtain the same phase diagram as suggested in
Ref.\ \cite{K00}. Now, the question is what changes in this picture if
$S$ becomes half-integer?
%

As it was mentioned before, the planar description becomes invalid
near the vortex core, where it becomes energetically favorable to lift
the spins from the easy plane in order to reduce the contribution of
the gradient terms in the action.  The vector $\vec{l}$ in the center
of a vortex is in fact perpendicular to the plane, and thus the
vortices present in the model are in fact not $Z$ vortices of the
purely $XY$ system, but rather $Z_{2}$ vortices of the Heisenberg
model. 
The topological charge of a vortex is $Q=-p\nu/2=\pm{1\over2}$,
where $p=\pm1$ is the sign of $\cos\theta$ in the vortex center, and
$\nu=\pm1$ is the usual $XY$ vorticity (the angle $\varphi$ changes by
$2\pi \nu$ when one goes around the center). Therefore, for
half-integer $S$ every vortex obtains the effective phase factor
$e^{i2\pi SQ}=e^{\pm i\pi/2}$, and after summation over possible
values of $Q$ the contribution of vortices vanishes. This has the
usual consequence of suppressing the Kosterlitz-Thouless
transition, which disappears for half-integer $S$ also for small $j$,
together with the corresponding gapped (Haldane)
phase. \cite{reviews2}  Thus one may expect that only the chiral gapless
phase should survive for half-integer $S$, in accordance with the
numerical results. \cite{Hik+01}

\begin{wrapfigure}{r}{66mm}
\epsfxsize=65mm
\centerline{\epsfbox{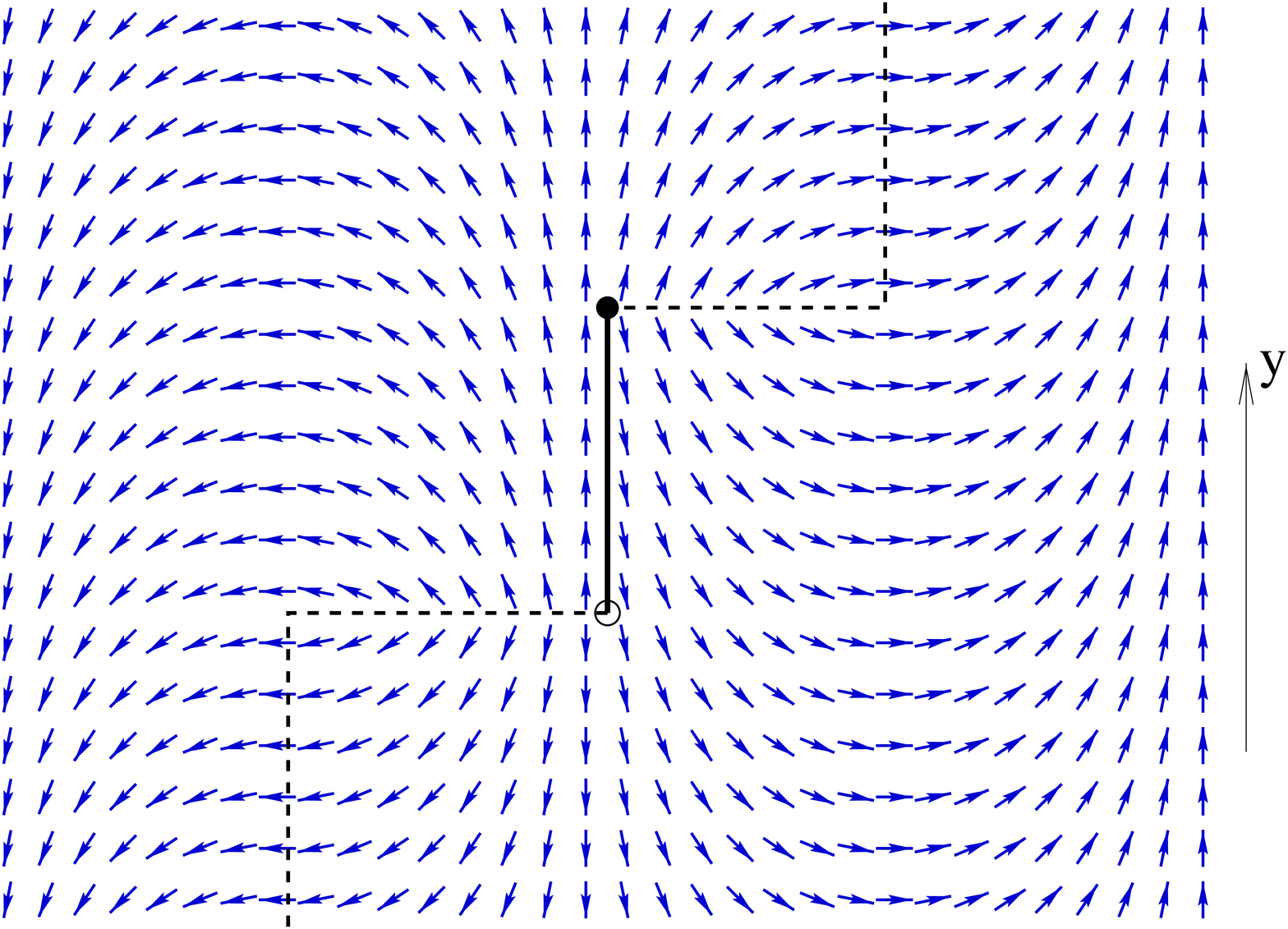}}
\caption{\label{fig:hiDW} An example of configuration with
discontinuity which may contribute to the destabilization of the
chiral gapless phase for half-integer $S$, see the text. The dashed
line shows the position of the chiral domain wall, and the
discontinuity is indicated by the solid line. The discontinuity is a
chiral domain wall with an additional $\pi$ phase jump; at both ends
of the discontinuity there are ``half-vortices'' which carry the
topological charge $Q=\pm {1\over4}$.} 
\end{wrapfigure}

Another, somewhat unexpected, effect of the topological term is that
it suppresses the fluctuations of the chiral domain
walls as well. Indeed, since every elementary fluctuation (a ``jump'' by
$\pi/\lambda_{0}$ in space) contains a ``bound vortex'', the
contribution of configurations with $n_{i}\not=0$ in (\ref{freedw1})
should  vanish. That means that the mechanisms of destabilizing the
chiral gapless phase for integer and half-integer $S$ must be very
different. At present, I am not able to suggest any efficient
mechanism for destroying the chiral phase for half-integer $S$. One
may speculate that high-energy  configurations with discontinuities 
may play some role 
(an example of such configuration is shown in Fig.\
\ref{fig:hiDW}). 
Another possibility would be that due to interaction between bound
vortices the energy of configurations describing the domain wall
fluctuations with even $n_{i}$ will be the lowest when all $n_{i}$
``bound vortices'' have the same topological charge $Q=+{1\over2}$ or
$-{1\over2}$ (note that a uniform sequence of $Q$ means an alternating
sequence of $p=\mbox{sign}{\cos\theta}$, since the vorticities $\nu$
{\em must} alternate). In any case, such explanations would mean a
strong increase of the Ising transition temperature in comparison to
the integer $S$ case.

%
%
%
%
%

{\em Acknowledgments. ---}
I would like to thank P. Azaria, T. Hikihara, T. Jolicoeur, C. Lhuillier, and
P. Lecheminant for the fruitful discussions on the subject. This work
was partly supported by Volkswagen-Stiftung through Grant I/75895.

\end{document}